\documentclass[twocolumn,tighten]{aastex63}
\usepackage{amssymb, amsmath,framed}

\usepackage{bm}
\expandafter\ifx\csname package@font\endcsname\relax\else
 \expandafter\expandafter
 \expandafter\usepackage
 \expandafter\expandafter
 \expandafter{\csname package@font\endcsname}
\fi
\def\bq{\begin{equation}}
\def\eq{\end{equation}}
\def\bqy{\begin{eqnarray}}
\def\eqy{\end{eqnarray}}

\def\p{\partial}

\def\p{\partial}

\def\R{\mathbb{R}}

 \def\q#1#2{q^{\hspace{1pt}#1}_{\,,\hspace{1pt}#2}}
 \def\a#1#2{a^{\hspace{1pt}#1}_{\,,\hspace{1pt}#2}}
 \def\d#1#2{\delta^{\hspace{1pt}#1}_{\,,\hspace{1pt}#2}}

\begin{document}
\title{\large{Potential for Liquid Water Biochemistry Deep under the Surfaces of the Moon, Mars, and beyond}}

\correspondingauthor{Manasvi Lingam}
\email{mlingam@fit.edu}

\author{Manasvi Lingam}
\affiliation{Department of Aerospace, Physics and Space Sciences, Florida Institute of Technology, Melbourne, FL 32901, USA}
\affiliation{Institute for Theory and Computation, Harvard University, Cambridge, MA 02138, USA}

\author{Abraham Loeb}
\affiliation{Institute for Theory and Computation, Harvard University, Cambridge, MA 02138, USA}

\begin{abstract}
We investigate the prospects for the past or current existence of habitable conditions deep underneath the surfaces of the Moon and Mars, as well as generic bound and free-floating extrasolar rocky objects. We construct a simple model that takes into account the thermal limits of life as well as the size, surface temperature, and relative radionuclide abundance of a given object and yields the spatial extent of the subsurface habitable region. We also investigate the constraint imposed by pressure on habitability, and show that it is unlikely to rule out the prospects for life altogether. We estimate the maximum biomass that might be sustainable in deep subsurface environments as a function of the aforementioned parameters from an energetic perspective. We find that it might be a few percent that of Earth's subsurface biosphere, and three orders of magnitude smaller than Earth's global biomass, under ideal circumstances. We conclude with a brief exposition of the prevalence of rocky objects with deep biospheres and methods for detecting signatures of biological activity through forthcoming missions to visit the Moon and Mars.\\
\end{abstract}

\section{Introduction} \label{SecIntro}
The habitable zone (HZ), in its classical form, delineates the region around the host star where liquid water could exist on the surface of a rocky planet \citep{KWR93}. The concept of the HZ has been extended in myriad directions by including new greenhouse gases, taking surface geology into account, expanding it to the galactic level, and many more \citep{Ram18}, thereby making it widespread in astrobiology, especially from a practical standpoint in the search for biosignatures. It is worth emphasizing, however that the canonical HZ \citep{KWR93} definition pertains to remote detection of habitability. Worlds outside the HZ may very well be habitable, as illustrated by the classic examples of Europa, Enceladus, and Titan in our Solar system. Another notable category of potentially habitable worlds is composed of free-floating planets and moons, which might either host liquid water on the surface \citep{Ste99,LL20} or comprise oceans beneath icy envelopes \citep{AS11}.

In a similar vein, one may investigate the habitability of rocky environments \emph{underneath} the surface. On Earth, subsurface rock-based (endolithic) environments are known to host rich and thriving biospheres \citep{Gold,EBC12}, and the total biomass in these settings is predicted to be $\sim 10\%$ of Earth's overall value \citep{WCW98,BPM18}. Hence, it is not surprising that a number of studies have investigated Martian subsurface habitability \citep{BIM92,Co14,MOM18,SKC19,CBM20}. However, \emph{generic} habitability studies of deep terrestrial biospheres are lacking, with the exception of \citet{MOP13}.

Our work differs from the aforementioned publication in the following respects: (i) we explicitly account for the possibility of free-floating rocky objects with only internal heating, (ii) we take other constraints (e.g., pressure and energy) into consideration, and (iii) we span a larger parameter space of physical variables. We calculate the extent of the habitable subsurface region in Sec. \ref{SecHabDep}, tackle other properties of deep biospheres in Sec. \ref{SecAdCon}, and summarize our findings in Sec. \ref{SecConc}.

\section{Spatial extent of subsurface biospheres}\label{SecHabDep}

\begin{figure*}
$$
\begin{array}{cc}
  \includegraphics[width=9.0cm]{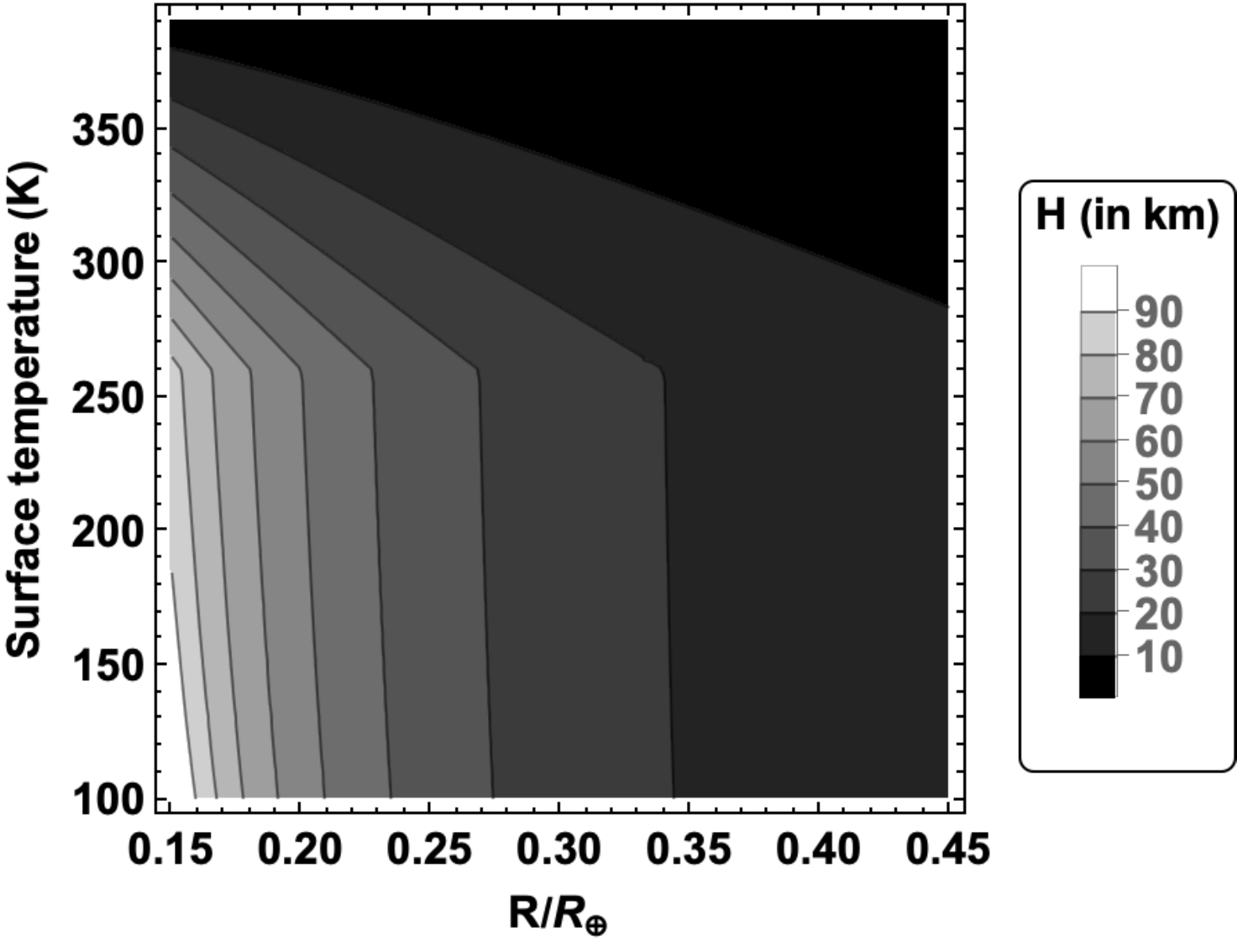} &  \includegraphics[width=8.9cm]{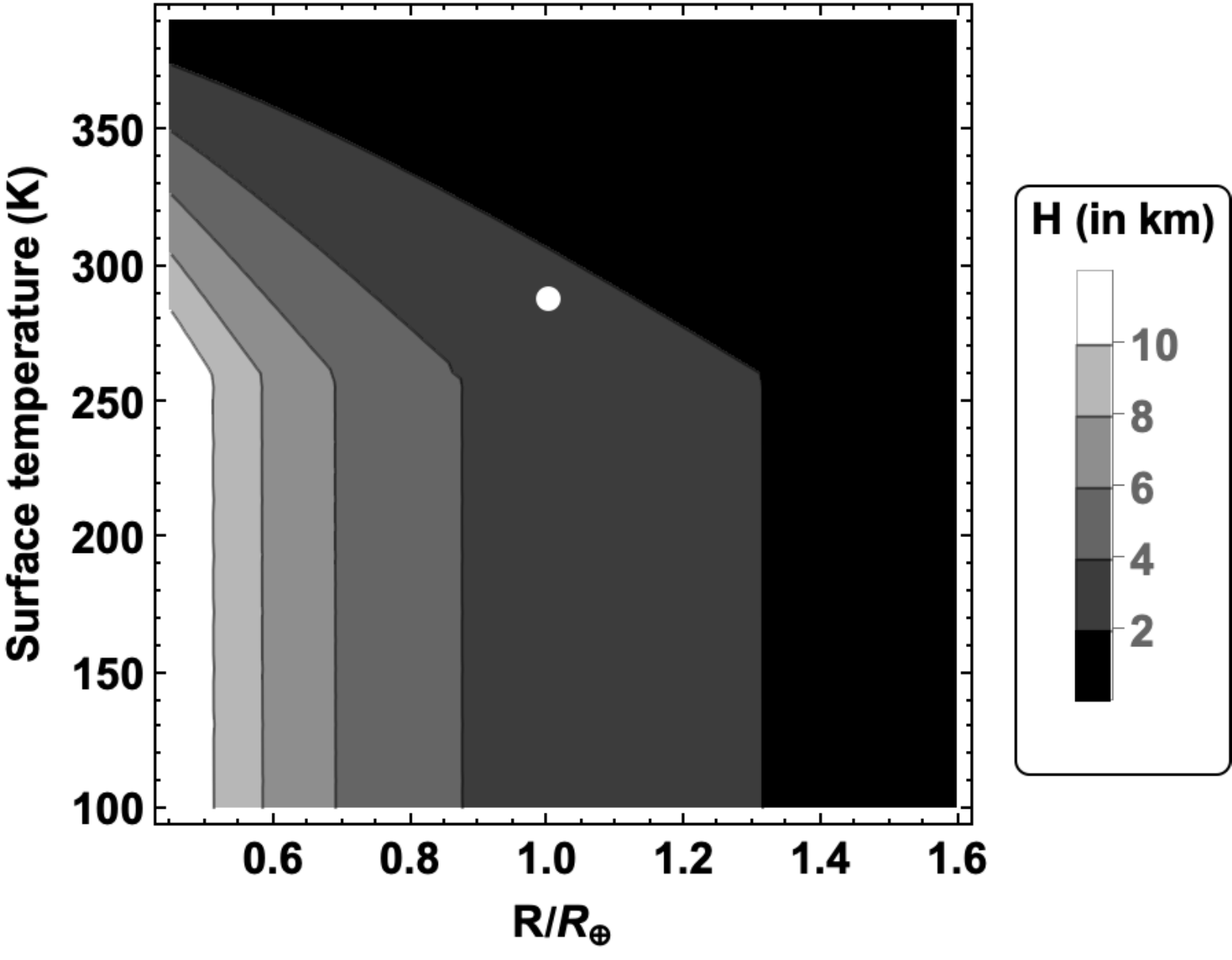}\\
\end{array}
$$
\caption{In both panels, the contour plots for the extent of the subsurface habitable region $H$ (in km) are plotted as functions of the surface temperature and radius of the object. In the right panel, the white circle encapsulates the Earth's parameters.}
\label{FigHZProp}
\end{figure*}

The minimum temperature for liquid water is roughly between $250$-$270$ K when the pressure is $\lesssim 1$ GPa \citep{CG07}. The cryophile \emph{Planococcus halocryophilus} grows at $258$ K \citep{MAB19}, which falls in this regime. Both experiments and theory indicate that lower temperatures would result in the vitrification of cells. The maximum temperature for liquid water's existence is sensitive to total pressure and exceeds $373$ K when the pressure is $> 1$ bar. However, at sufficiently high temperatures, a number of inexorable biochemical and biophysical constraints come into play \citep{Cla14,McK14}. The hyperthermophile \emph{Methanopyrus kandleri} survives at $395$ K \citep{TNT08}, although the limits for organic life might extend up to $\sim 450$ K \citep{BXY15}. Based on the preceding considerations, $T_\mathrm{min} = 260$ K and $T_\mathrm{max} = 400$ K demarcate our thermal range.

In its most general form, the heat-transfer equation is expressible as \citep[Equation 4.13]{Mel11}:
\begin{equation}\label{HTEIn}
\frac{\partial T}{\partial t} = \nabla \cdot \left(\kappa \nabla T \right) + \frac{\mathcal{H}}{C_P},
\end{equation}
where $T$ is the temperature, whereas $\kappa$, $C_P$ and $\mathcal{H}$ denote the thermal diffusivity, specific heat capacity at constant pressure, and the rate of heating per unit mass. As our goal is to develop a simple model, we will follow the approach delineated in \citet{LL19}, which generalizes the slab models described in \citet{AS11} and \citet{MOP13}. 

To begin with, we tackle the steady-state solution; although $\mathcal{H}$ is time-dependent, it is subject to significant change only over long timescales of several Gyr \citep[Section 6.24]{TS02}. We suppose that Fourier's law is valid, thereby yielding
\begin{equation} \label{Fourier}
    \mathcal{Q} +  k \frac{d T}{d r} = 0, 
\end{equation}
where $\mathcal{Q} \equiv \mathcal{Q}(r)$ is the heat flux at $r$ and $k \equiv k(T)$ denotes the thermal conductivity of the rock(s). The former quantity is modeled by using
\begin{equation} \label{Hflux}
    \mathcal{Q} = \frac{Q}{4\pi r^2} \times \frac{4\pi r^3/3}{4\pi R^3/3} = \frac{Q}{4\pi R^2} \frac{r}{R},
\end{equation}
where $R$ is the radius of the world, while $Q$ embodies the total internal heat flow. Therefore, the above ansatz posits that the heat flux is the ratio of the heat flow at this location ($Q_E$) and the area encompassed ($4\pi r^2$). The former variable is determined by assuming that the heating is uniformly distributed in the interior, amounting to $Q_E \approx Q \times \left(\frac{4\pi}{3}r^3/\frac{4\pi}{3} R^3\right)$. Furthermore, we specify $Q$ following the standard prescription \citep[e.g.,][]{VC09,LL21}:
\begin{equation} \label{Qans}
    Q = \Gamma\, Q_\oplus \left(\frac{M}{M_\oplus}\right),
\end{equation}
where $Q_\oplus \approx 4.4 \times 10^{13}$ W is the Earth's heat flow \citep{Kam11}, $\Gamma$ is the abundance of long-lived radionuclides per unit mass measured relative to Earth, and $M$ is the object's mass, which is related to $R$ via the scaling $M \propto R^{3.7}$ \citep{ZSJ16}. Note that (\ref{Fourier}) and (\ref{Hflux}) ensure that the geothermal flux condition is satisfied at the surface, and $dT/dr = 0$ at $r = 0$.

The remaining component is $k(T)$. Evidently, this quantity is subject to much variability as it depends upon the characteristics of the rocks (e.g., sedimentary, metamorphic). For instance, the thermal conductivity of sandstone is $\sim 2$ times higher than basalt and granite at $273$ K \citep{HTK16}. This issue is compounded by the absence of empirical studies of heat conductivity at very low temperatures, which has necessitated reliance on heuristic models. Bearing these caveats in mind, we will consider a rocky object with a Mars-like composition, whose thermal conductivity at the surface is $\lesssim 2$ times smaller than that of Earth \citep[cf.][]{WHN09,PJM17}. For this model, we treat the crust as being dominated by basalt and employ the relationship \citep[Equation 1]{CLH10}:
\begin{equation}
    k(T) \approx \mathcal{A} + \frac{\mathcal{B}}{T},
\end{equation}
where $\mathcal{A} = 0.4685$ W m$^{-1}$ K$^{-1}$ and $\mathcal{B} = 488.19$ W m$^{-1}$. After integrating (\ref{Fourier}) and drawing upon the preceding formulae, we arrive at
\begin{equation}\label{GenSol}
    \frac{Q \left(R^2 - r^2\right)}{8\pi R^3} = \mathcal{A} \left(T - T_s\right) + \mathcal{B} \ln\left(\frac{T}{T_s}\right),
\end{equation}
where $T_s$ is the average surface temperature.\footnote{Although our solution of the heat-transfer equation is time-independent, we point out that real-world objects (e.g., Moon) exhibit substantive temporal and spatial variations in $T_s$. Hence, $T_s$ should be envisioned as the spatially and temporally averaged value of the surface temperature.} We define $F(T) \equiv \mathcal{A} \left(T - T_s\right) + \mathcal{B} \ln(T/T_s)$, which plays a prominent role in the subsequent analysis.

There are two distinct cases that we must investigate in order to determine the spatial extent $H$ of the putative subsurface biosphere. In the first scenario, we have $T_s > T_\mathrm{min}$ so that the conditions are warm enough in principle for surface-based life to survive. In this case, as explained in Appendix \ref{AppA}, we find that
\begin{equation}\label{Hreg1}
    \frac{H}{R} = 1 - \sqrt{1 - \frac{8\pi R}{Q} F(T_\mathrm{max})},
\end{equation}
where $F(T)$ was introduced below (\ref{GenSol}). For $H \ll R$, simplifying (\ref{Hreg1}) further leads to
\begin{equation}\label{Hreg1sim}
    H \approx \frac{4\pi R^2 F(T_\mathrm{max})}{Q}.
\end{equation}

\begin{figure}
\includegraphics[width=7.0cm]{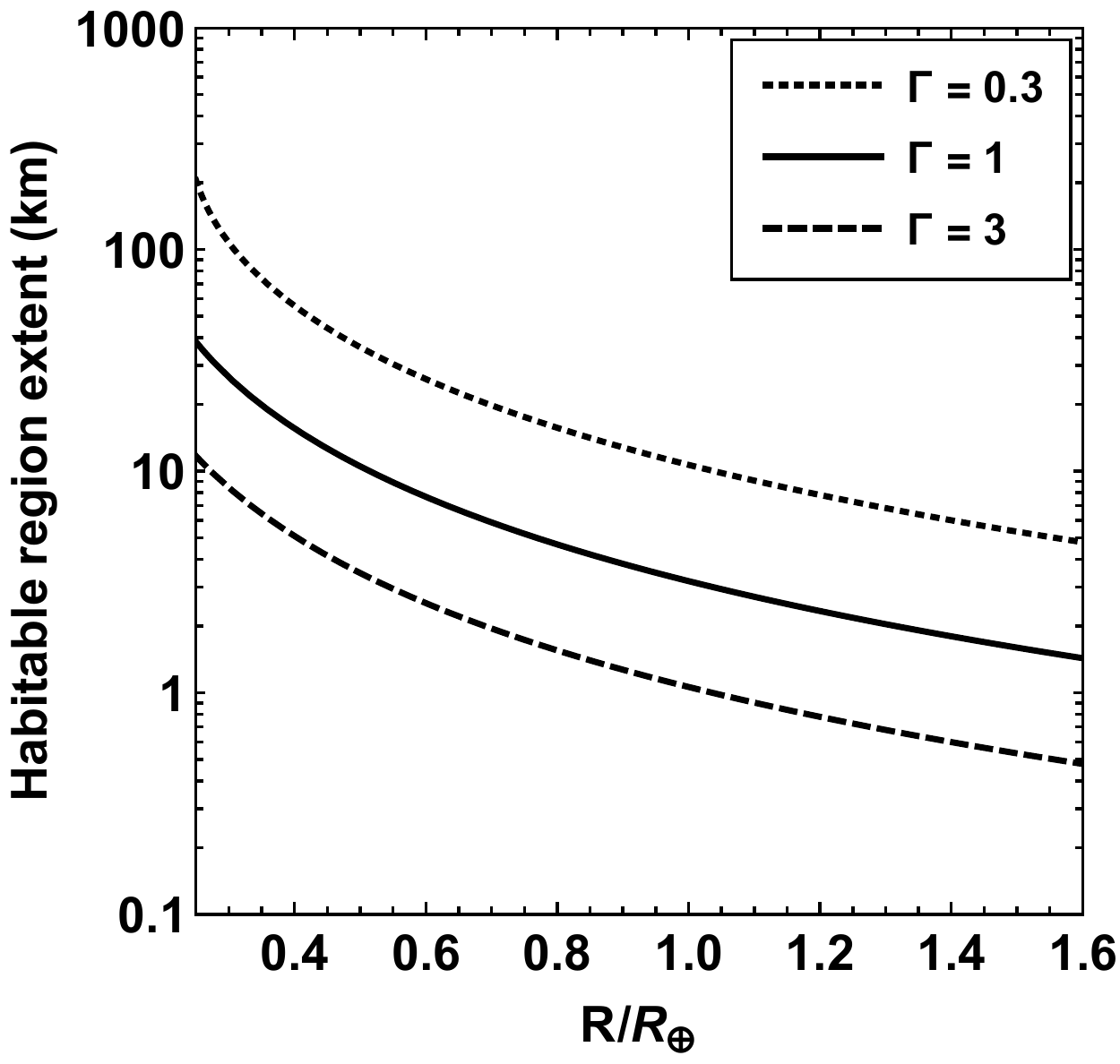} \\
\caption{Extent of the habitable region as a function of the radius of the object for differing relative radionuclide abundance. This plot pertains specifically to free-floating rocky objects.}
\label{FigHZFF1}
\end{figure}

In the second case, we consider $T_s < T_\mathrm{min}$, implying that life near the surface is not feasible. In this setting, $H$ is computed in Appendix \ref{AppA} and found to be
\begin{equation}\label{Hreg2}
\frac{H}{R} = \sqrt{1 - \frac{8\pi R}{Q} F(T_\mathrm{min})} - \sqrt{1 - \frac{8\pi R}{Q} F(T_\mathrm{max})}.
\end{equation}
If the lower and upper limits of this biosphere, delineated in Appendix \ref{AppA}, are much smaller than $R$, the above equation simplifies to
\begin{equation}\label{Hreg2sim}
    H \approx \frac{4\pi R^2}{Q} \left[F(T_\mathrm{max}) - F(T_\mathrm{min})\right].
\end{equation}
For the second case, we note that free-floating rocky objects would fall under this category. The surface temperature is purely set by internal heat and is determined via the Stefan-Boltzmann law, which yields
\begin{equation}\label{TradH}
T_s \approx 35\,\mathrm{K}\,\,\Gamma^{1/4}\left(\frac{R}{R_\oplus}\right)^{0.43}.
\end{equation}
From inspecting (\ref{Hreg1}) and (\ref{Hreg2}), we notice that $H$ is real-valued only if the following criterion holds true:
\begin{equation}
    \frac{Q}{8\pi R} > F(T_\mathrm{max}).
\end{equation}
In what follows, we will implicitly presume that this condition is fulfilled. Broadly speaking, this translates to $R \gtrsim 0.1\,R_\oplus$, with the caveat that the exact magnitude is sensitive to $T_s$.

From the preceding discussion, evidently $H$ is a complex function of $\Gamma$, $R$ and $T_s$, except for free-floating rocky objects (see Appendix \ref{AppA}). In Fig. \ref{FigHZProp}, we have plotted $H$ as a function of $T_s$ and $R$ while fixing $\Gamma = 1$. Note that this figure corresponds to rocky objects where the surface temperature is set by either stellar or tidal heating and not by internal (i.e., radiogenic and primordial) heating. There are two broad trends that can be inferred. First, $H$ exhibits a fairly strong dependence on $R$, which becomes evident from (\ref{Hreg1sim}) and (\ref{Hreg2sim}) as well. Second, $H$ is nearly independent of $T_s$ for $T_s < T_\mathrm{min}$, viz., $H$ is largely unaffected by the properties of the region between $T = T_s$ and $T = T_\mathrm{min}$. 

As we chose a Mars-like composition, we compare our results against prior studies of the Martian subsurface biosphere. From Appendix \ref{AppA}, we ascertain that the habitable subsurface region spans $H_1 \approx 4.2$ km and $H_2 \approx 13.6$ km beneath the surface, thereby corresponding to $H = 9.4$ km. Note that $H_1$ and $H_2$ signify the locations of the top ($T = T_\mathrm{min}$) and bottom ($T = T_\mathrm{max}$) layers of the subsurface habitable region, as measured from the object's surface. This result is consistent with the slab model of \citet{MOP13}, where it was determined that the biosphere might span $H_1 \approx 4.5$ to $H_2 \approx 14.0$ km, thus yielding $H = 9.5$ km. In a more detailed analysis, \citet{CLH10} estimated $H \approx 2.3$-$4.7$ km at the equator and $H \approx 6.5$-$12.5$ km at the poles. If one takes the arithmetic mean of these two extremes, we end up with $H \approx 4.4$-$8.6$ km; the upper bound is therefore not far removed from our model's predictions.

Next, we turn our attention to free-floating objects where $H$ is only a function of $R$ and $\Gamma$ as outlined in Appendix \ref{AppA}. We have plotted the ensuing behavior of $H$ in Fig. \ref{FigHZFF1} for different choices of $\Gamma$. Aside from the expected sensitivity to $R$, we see that $\Gamma$ plays a fairly important role in regulating the spatial extent of the habitable region. An interesting point worth mentioning is that higher $\Gamma$ augurs well for surface habitability in some respects \citep{LL20}, but it suppresses the extent of the subsurface habitable zone and other characteristics encountered hereafter because of the inverse dependence on $\Gamma$ in (\ref{Hreg1sim}) and (\ref{Hreg2sim}), after using the scaling for $Q$ prescribed in (\ref{Qans}).

\section{Additional limits on subsurface biospheres}\label{SecAdCon}
In this Section, we sketch other constraints on the characteristics of putative subsurface biospheres.

\begin{figure}
\includegraphics[width=7.0cm]{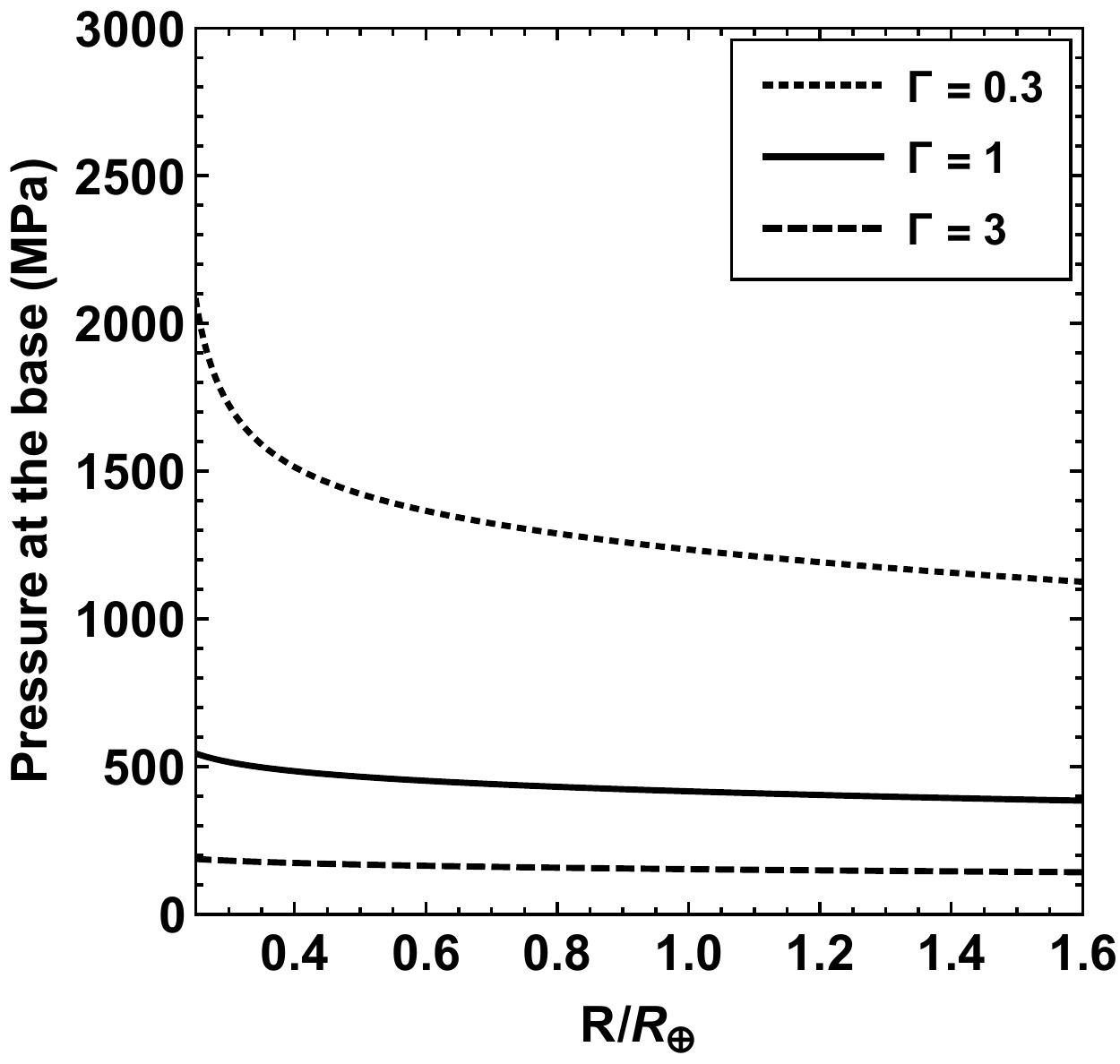} \\
\caption{The pressure at the base of the subsurface habitable region (in MPa) as a function of the radius of the world. This plot pertains specifically to free-floating rocky objects.}
\label{FigEnPres}
\end{figure}

\subsection{The role of pressure}
With recent advances in high-pressure geomicrobiology, there are grounds for supposing that high pressure may stymie the prospects for life \citep{PD13}. Hence, it is worth calculating the pressure at the base where $T = T_\mathrm{max}$ is attained, and compare it against the known limits for microbes on Earth. The pressure ($P_b$) at this location is
\begin{equation}\label{Pbdef}
  P_b \approx \rho_c g_\oplus \left(\frac{R}{R_\oplus}\right)^{1.7} H_2, 
\end{equation}
where $\rho_c \approx 2.7 \times 10^3$ kg/m$^3$ is the roughly constant density of the crust \citep{WHN09}, $g_\oplus$ is the Earth's surface gravity and $H_2$ is the lower layer depth of the subsurface habitable region defined in Appendix \ref{AppA}; we have also made use of $g \propto M/R^2$. If we presume that $H_2 \ll R$, we can make use of (\ref{Hreg1sim}) to obtain
\begin{equation}\label{Pbsimp}
    P_b \approx \frac{4\pi \rho_c g_\oplus R_\oplus^2 F(T_\mathrm{max})}{\Gamma Q_\oplus}.
\end{equation}
An unusual consequence of this equation is the absence of any \emph{explicit} dependence on $R$; in other words, to leading order $P_b$ depends only on $\Gamma$ and $T_s$; note that the latter has a weak dependence on $R$ as seen from (\ref{TradH}).

In principle, one can calculate $P_b$ as a function of $T_s$, $\Gamma$ and $R$. However, we will focus exclusively on free-floating rocky objects. The reason for doing so is that $H_2$ in (\ref{Pbdef}) is predicted to be larger, implying that the value of $P_b$ thus obtained constitutes an upper bound for fixed $R$ and $\Gamma$. To offer an example, our model yields $P_b \approx 120$ MPa for Mars, whereas a free-floating Mars-sized world would have $P_b \approx 460$ MPa; in both cases, we have set $\Gamma = 1$. Fig. \ref{FigEnPres} depicts $P_b$ as a function of $R$ and $\Gamma$. It is apparent that the pressure becomes nearly independent of the radius in accordance with (\ref{Pbsimp}). Moreover, for $\Gamma = 1$, we find that $P_b < 1$ GPa holds true for rocky objects with sizes comparable or larger than the Moon.

On Earth, the piezophile \emph{Thermococcus piezophilus} is known to survive at pressures of roughly $125$ MPa \citep{MAB19}, which is comparable to the values in Fig. \ref{FigEnPres} for $\Gamma \gtrsim 1$. Furthermore, laboratory experiments entailing \emph{Escherichia coli} have found that these bacteria rapidly acquire piezoresistance in the GPa range \citep{VMB11}. Hence, our analysis suggests that the constraints imposed by pressure are insufficient to rule out the prospects for life altogether. One caveat worth noting here is that organisms in deep subsurface environments would need to concomitantly withstand multiple extremes, but our knowledge of polyextremophiles is filled with lacunae \citep{HGT13}.

\subsection{Biomass in the deep biosphere}
Ever since the pioneering study by \citet{Gold}, many studies have sought to quantify the biomass embedded in Earth's subsurface biosphere. The estimates vary by more than an order of magnitude due to the variety of methods deployed for this calculation, as well as the inherent uncertainties surrounding some of the salient parameters \citep{MLD18}. 

A common method for calculating the biomass ($M_\mathrm{bio}$) utilizes the scaling \citep{Gold,WCW98}:
\begin{equation}\label{Mbio}
   M_\mathrm{bio} \propto \frac{\mathcal{C}_\mathrm{cell}\,\delta_s\,  \phi_\mathrm{cell}\,\mathcal{V}_s}{\mathcal{V}_\mathrm{cell}},
\end{equation}
where $\delta_s$ and $\mathcal{V}_s$ represent the porosity and volume of the subsurface habitable region, $\phi_\mathrm{cell}$ denotes the fraction of the pore space occupied by microbes, while $\mathcal{C}_\mathrm{cell}$ and $\mathcal{V}_\mathrm{cell}$ are the typical organic carbon mass and volume of a typical cell, respectively. It is evident that most of the variables are either geological or biological in nature, and therefore are subject to considerable variability; even on Earth, these parameters are not tightly constrained. The sole exception is $\mathcal{V}_s$ because it is possible to loosely estimate it using the results from our model.

Note that $\mathcal{V}_s \propto R^2 H$ as it quantifies the area of a spherical shell with thickness $H$. Although $H$ is a complex multivariable function, we can draw upon the preceding results. For $T_s < T_\mathrm{min}$ - which is true for most rocky objects of interest herein - it was shown in Fig. \ref{FigHZProp} that $H$ becomes effectively independent of $T_s$. This leaves us with the parameters $R$ and $\Gamma$. From (\ref{Hreg1sim}) and (\ref{Hreg2sim}), which are fairly accurate for $R \gtrsim 0.1\,R_\oplus$, we see that the same scaling applies, namely, $H \propto R^2/Q$.\footnote{In fact, this trend holds true irrespective of the exact form of $k$ because the only difference is that $F(T)$ would have to be replaced by another function.} Hence, from these scalings, we find that $\mathcal{V}_s \propto R^4/Q \propto R^{0.3}/\Gamma$ with the last relationship following from (\ref{Qans}). 

Thus, with all other factors in (\ref{Mbio}) held fixed apart from $\mathcal{V}_s$, the biomass is estimated as
\begin{equation}\label{Mscal}
    M_\mathrm{bio} \sim 20\,\mathrm{Pg\,C}\,\left(\frac{1}{\Gamma}\right) \left(\frac{R}{R_\oplus}\right)^{0.3},
\end{equation}
where the normalization was adopted from the mean value of \citet[pg. 712]{MLD18}. A striking aspect of (\ref{Mscal}) is the relatively weak dependence on $R$. In principle, a Super-Earth with $R \approx 1.5\,R_\oplus$ and $\Gamma \approx 0.2$ might support a subsurface biosphere whose biomass is $\sim 20\%$ that of Earth's total biomass, which equals $550$ Pg C \citep{BPM18}. In actuality, however, we caution that $M_\mathrm{bio}$ is constrained by energy, nutrient, and miscellaneous limitations, owing to which $M_\mathrm{bio}$ constitutes an idealized loose upper bound.

Neither energy nor nutrient limitation are easy to model, and require a case-by-case treatment.Instead, we will focus on a single energetic pathway. Among the various energy sources, radiolysis has garnered widespread attention \citep{LHL05,SVL07,Atri,DSD18}, especially after the discovery of \emph{Desulforudis audaxviator}, a sulfate-reducing bacterium living $2.8$ km beneath Earth's surface that derives its energy from this source \citep{CBA08}. The \emph{maximum} subsurface biomass ($M_\mathrm{max}$) that might be supported by methanogenesis with the reactant (H$_2$) derived from radiolysis is
\begin{equation}
  M_\mathrm{max} \approx \frac{\mathcal{C}_\mathrm{cell}\, \Delta G\, \bar{\mathcal{P}}\, \mathcal{V}_s }{P_\mathrm{bio}},  
\end{equation}
where $\bar{\mathcal{P}}$ is the average volumetric production rate of H$_2$ via radiolysis, $\Delta G$ is the Gibbs free energy of methanogenesis, and $P_\mathrm{bio}$ is the minimal power required for the viability of a single microbe. As most of these parameters are unknown for other objects, we hold them fixed at the average values on Mars and/or Earth. In the Noachian era, $\bar{\mathcal{P}}$ was probably $\lesssim 10$ times higher than today due to the higher abundance of radionuclides \citep{TMS18}, while the other parameters were ostensibly similar. Thus, by this reckoning, the total H$_2$ that might arise from radiolysis on modern Mars is roughly an order of magnitude lower than the Noachian mean value of $\sim 3 \times 10^{10}$ moles/year \citep{TMS18}. It was estimated by \citet{SKC19} that a total H$_2$ flux of $\sim 1.3 \times 10^8$ moles/year could support $\sim 10^{27}$ cells. By employing these predictions along with $\mathcal{C}_\mathrm{cell} \sim 2.1 \times 10^{-17}$ kg \citep[pg. 707]{MLD18}, $\bar{\mathcal{P}} \propto \Gamma$ (because the production rate is proportional to radionuclide concentration) and the scaling $\mathcal{V}_s \propto R^{0.3}/\Gamma$ derived earlier, we arrive at
\begin{equation}\label{Mmaxcal}
  M_\mathrm{max} \sim 0.6\,\mathrm{Pg\,C}\,\left(\frac{R}{R_\oplus}\right)^{0.3}.
\end{equation}

Upon comparing (\ref{Mmaxcal}) and (\ref{Mscal}), we find that $M_\mathrm{max}$ is $\sim 3\%$ of Earth's subsurface biosphere for $R \approx R_\oplus$. Moreover, at this radius, $M_\mathrm{max}$ is merely $\sim 10^{-3}$ that of Earth's global biomass \citep{BPM18}. The weak scaling in (\ref{Mmaxcal}) implies that $M_\mathrm{max}$ is only $\lesssim 2$ times smaller on objects such as the Moon ($R = 0.27\,R_\oplus$) and Mars ($R = 0.53\,R_\oplus$). Hence, in principle, both these objects might have been capable, at some point perhaps, of hosting deep biospheres that were approximately three orders of magnitude smaller than Earth's global biomass.

In closing, we reiterate that only energetic and thermal (which dictates $\mathcal{V}_s$) constraints were considered. Aside from limitations imposed by nutrients and reactants, liquid water is obviously another vital requirement. However, this issue is hard to quantify because it depends, \emph{inter alia}, on the pathways of object formation, volatile delivery by impactors, crust composition, and potentially the deep water cycle \citep{PSB17}. It is estimated that Earth's crust comprises $\sim 4 \times 10^{23}$ mL of water \citep[Table 1]{PSB17}. If roughly $\sim 10\%$ of water is assumed to be present in the subsurface biosphere region and it hosts $\sim 10^{30}$ cells, we find that the average cell density is $\sim 2.5 \times 10^7$ cells/mL; this estimate is compatible with certain sophisticated models \citep{MLD18}. Hence, provided that other rocky objects are not significantly devoid of volatiles, it seems unlikely \emph{prima facie} that subsurface biospheres would be limited by water availability.

\section{Conclusion}\label{SecConc}
In this work, we examined the prospects for life in deep terrestrial environments. While this subject has been investigated for specific rocky objects (e.g., Earth), we constructed simple models to assess the salient features of such putative biospheres. 

We began by studying the extent of the habitable region ($H$) associated with deep biospheres as a function of the surface temperature, relative radionuclide abundance ($\Gamma$) and size ($R$) of the world. We showed that $H$ may span orders of magnitude for different combinations of these parameters. For rocky objects that are Moon-sized and larger, we determined that the scaling $H \propto R^{-1.7} \Gamma^{-1}$ is valid. We compared the results from our model with prior studies (involving Mars) and found that they exhibit good agreement. Then we investigated two other constraints on the habitability of these settings. First, we studied the pressure at the base of the biosphere and showed that it is nearly independent of $R$. We determined that pressure is unlikely to completely rule out the prospects for putative lifeforms. Lastly, we presented simple estimates for the total biomass that could be supported in deep terrestrial habitats. We determined that this biomass may depend weakly on $R$, and that it might be a few percent of Earth's deep biosphere and three orders of magnitude smaller than Earth's global biomass.

In closing, two factors merit highlighting. First, the number of rocky planets in the HZs of host stars is $\lesssim 0.3$ per star \citep{ZH19}. Even if not all of these objects are capable of sustaining liquid water on the surface over long timescales (e.g., Mars), they may still have the capacity to support deep biospheres. Furthermore, gravitational microlensing studies \citep{SBM12,BDG19} and extrapolation of the size distribution function of interstellar objects \citep{SL19,RL19} both indicate that the number of free-floating objects with sizes larger than the Moon is $\gtrsim 10$-$100$ per host star.\footnote{The Moon is chosen as the lower threshold because our models are rendered valid in this domain.} Even if only a small fraction of them were formed inside the snow line (prior to ejection) and are predominantly rocky, they may nevertheless outnumber planets in the HZs of stars. 

Second, because deep biospheres are situated underneath the surface, detecting unambiguous signatures of biological activity is not readily feasible via remote sensing techniques. The most obvious solution is to carry out \emph{in situ} studies of rocky objects in our backyard such as the Moon and Mars. The Moon was habitable shortly after its formation \citep{SMC18}, and it is not altogether inconceivable that some traces and markers of life might survive to this day. NASA's Artemis program can pave the way for lunar subsurface exploration via establishing a sustainable base on the Moon purportedly by 2024.\footnote{\url{https://www.nasa.gov/specials/artemis/}} 

While methods like transient electromagnetic sounding can reveal the presence of subsurface water, advances in drilling technologies are necessary to develop autonomous machines capable of reaching km-scale depths \citep{CBM20}. Alternatively, one may carry out excavations at the sites of large craters on Mars to extract and analyze materials derived from the subsurface \citep{CB02}. A more ambitious (and technically feasible) enterprise entails combing the Oort cloud for the nearest Moon-sized extrasolar objects and sending fast spacecraft with appropriate payloads to scrutinize them \citep{HEH20}.

\acknowledgments
This work was supported by the Breakthrough Prize Foundation, Harvard University's Faculty of Arts and Sciences, and the Institute for Theory and Computation (ITC) at Harvard University.

\appendix
\section{The spatial extent of the subsurface habitable region}\label{AppA}
As noted in Sec. \ref{SecHabDep}, there are two cases that we need to consider in order to calculate the spatial extent of the subsurface habitable region ($H$). We examine them in detail below.

In the first case, $T_s > T_\mathrm{min}$. In this context, the top layer of the habitable region starts at the surface, i.e., at $r = R$. Let us suppose that the bottom layer of the habitable region - where $T = T_\mathrm{max}$ is attained - occurs at a depth $H$ beneath the surface (namely $r = R - H$). In this particular scenario, therefore, the spatial extent of the habitable subsurface region is $H$ and is found by setting $T = T_\mathrm{max}$ and $r = R - H$ in (\ref{GenSol}) and using the definition of $F(T)$ introduced just below (\ref{GenSol}). After doing so, we end up with
\begin{equation}
  \frac{Q \left[R^2 - (R-H)^2\right]}{8\pi R^3} = F(T_\mathrm{max}),
\end{equation}
which can be further simplified to yield
\begin{equation}
    \frac{Q \left(2R H - H^2\right)}{8\pi R^3} = F(T_\mathrm{max}),
\end{equation}
which is a quadratic equation in $H$. By solving this equation for $H$ and eliminating the non-physical root (which exhibits $H > R$), we end up with (\ref{Hreg1}). 

Now, let us tackle the second setting wherein $T_s < T_\mathrm{min}$. In this case, the top layer of the habitable subsurface region will \emph{not} be at the surface, but will begin at a depth $H_1$ (namely $r = R - H_1$) below it. One can solve for $H_1$ as follows: at the location $r = R - H_1$, the temperature is equal to $T_\mathrm{min}$. By substituting this value of $r$ and $T$ in (\ref{GenSol}), and repeating the same procedure in the above paragraph, we find
\begin{equation}\label{H1}
    \frac{H_1}{R} = 1 - \sqrt{1 - \frac{8\pi R}{Q} F(T_\mathrm{min})}.
\end{equation}
Next, the bottom layer of the habitable surface region will occur when the temperature is equal to $T_\mathrm{max}$. If we denote the depth below the surface where this occurs by $H_2$ (corresponding to $r = R - H_2$), we can solve for the latter quantity by demanding that $T = T_\mathrm{max}$ at $r = R - H_2$. After making use of (\ref{GenSol}) and carrying out the resultant algebra, we end up with
\begin{equation}\label{H2}
    \frac{H_2}{R} = 1 - \sqrt{1 - \frac{8\pi R}{Q} F(T_\mathrm{max})}.
\end{equation}
In this scenario, recall that the habitable subsurface region begins at $H_1$ and is terminated at $H_2$. Therefore, the spatial extent of the subsurface habitable region is $H = H_2 - H_1$, which leads us to (\ref{Hreg2}).

For free-floating worlds, the second case is indeed valid. Here, let us recall that
\begin{equation}
 F(T_\mathrm{min}) = \mathcal{A} \left(T_\mathrm{min} - T_s\right) + \mathcal{B} \ln\left(\frac{T_\mathrm{min}}{T_s}\right),
\end{equation}
\begin{equation}
 F(T_\mathrm{max}) = \mathcal{A} \left(T_\mathrm{max} - T_s\right) + \mathcal{B} \ln\left(\frac{T_\mathrm{max}}{T_s}\right),
\end{equation}
where $T_s$ in both of the above equations is set by (\ref{TradH}). One can then substitute these two expressions into (\ref{Hreg2}) to solve for $H$ as a function of $R$ and $\Gamma$. Due to the cumbersome nature of the expression, it is not explicitly provided here.


\end{document}